\documentclass{webofc}
\usepackage[varg]{txfonts}   
\usepackage{graphicx}	
\usepackage{amsmath}	
\usepackage{amssymb}
\usepackage[T1]{fontenc}
\usepackage{microtype}

\newcommand{\ramses}{\textsc{ramses}}
\def\rv{r_{\rm 500}}

\begin{document}
\title{Internal dark matter structure of the most massive galaxy clusters since redshift 1}
%
%

\author{\firstname{Amandine M. C.} \lastname{Le Brun}\inst{1,2}\fnsep\thanks{\email{amandine.le-brun@obspm.fr} (AMCLB)} 
        \and
	\firstname{Romain} \lastname{Teyssier}\inst{3}
}

\institute{PSL Fellow, LUTh, Observatoire de Paris, PSL Research University, CNRS, Universit\'e de Paris, 92195 Meudon, France
\and
           Laboratoire AIM, IRFU/D\'epartement d'Astrophysique -- CEA/DRF -- CNRS -- Universit\'e de Paris, B\^at. 709, CEA-Saclay 91191 Gif-sur-Yvette Cedex, France
\and
           Institute for Computational Science, University of Z\"urich, CH-8057 Z\"urich, Switzerland
          }

\abstract{%
 We investigate the evolution of the dark matter density profiles of the most massive galaxy clusters in the Universe. Using a `zoom-in' procedure on a large suite of cosmological simulations of total comoving  volume of  $3\,(h^{-1}\,\rm Gpc)^3$,  we study the 25 most massive clusters in  four redshift slices from $z\sim 1$ to the present. The minimum mass  is $M_{500} > 5.5 \times 10^{14}$ M$_{\odot}$  at $z=1$. Each system has more than two million particles within $\rv$. Once scaled to the critical density at each redshift, the dark matter profiles within $\rv$ are strikingly similar from $z\sim1$ to the present day, exhibiting a low dispersion of 0.15 dex, and showing little evolution with redshift in the radial logarithmic slope and scatter. They have the running power law shape typical of the NFW-type profiles, and their inner structure,  resolved to $3.8\,h^{-1}$ comoving kpc at $z=1$, shows no signs of converging to an asymptotic slope. Our results suggest that this type of  profile is already in place at $z>1$ in the highest-mass haloes in the Universe, and that it remains exceptionally robust to merging activity. 
}
\maketitle
\section{Introduction}
\label{intro}

This proceeding is entirely based upon \cite{LeBrun2018} and $M_{\Delta}$ is always the mass within radius $r_{\Delta}$, the radius within which the mean mass density is $\Delta$ times the critical density at the cluster redshift. Even though the theoretical framework for the formation of large-scale structures in a dark-matter-dominated universe was first formalised over 40 years ago \citep[e.g.][]{Peebles1980},  observations have only recently conclusively confirmed the  conceptual paradigm within which it takes place  \citep[e.g.][]{WMAP9,Planck2013,Planck2015}. The hierarchical collapse of dark matter into `haloes' in the dark-energy dominated cold dark matter ($\Lambda$CDM) model is nowadays the basis of our understanding of hierarchical galaxy and structure formation. The most massive galaxy clusters hold a peerless place in this hierarchy. Being mostly dark matter-dominated, they have deep potential wells and thus gravity is the dominant mechanism driving their evolution. Hence, they are least affected by complex non-gravitational galaxy formation processes \citep[e.g.][]{Cui2014,Martizzi2014,Velliscig2014}. They are detectable up to high redshift, and complementary techniques can probe both their internal structure and global properties. Hence, they are perfect objects for testing these theories. The emergence of Sunyaev-Zeldovich (SZ) surveys gave rise to a burst in the number of high-redshift clusters \citep{PSZ1,has13,Bleem2015,PSZ2}. Many of these are high mass, and the feasibility of obtaining high-quality structural and scaling information from these objects has now been proven \citep[e.g.][]{Bartalucci2017,sch16}.  In particular, X-ray observations of these bright systems can probe the $[0.05-1]\, \rv$ radial range quite easily \citep[e.g.][]{Bartalucci2017}. Numerical simulations of structure formation in the $\Lambda$CDM cosmology make testable predictions, such as the presence of a quasi-universal, cuspy, dark matter density profile \citep{Navarro1997}. However, existing simulations are ill-suited for studying high-mass, high redshift systems, as oft their resolution is scant, and/or simulated objects are in short supply. This study extends existing works on the shape of massive dark matter haloes  \citep{Tasitsiomi2004,Gao2012,Wu2013} by augmenting the spatial or mass resolution and the amount of systems. More essentially, we simulate the most massive galaxy clusters at $z>0$, and not just the progenitors of the $z=0$ ones. 

\section{Simulations and data processing}
\label{sec:sims}

We moulded our simulations to assemble a fairly-big sample ($\sim 50$) of massive ($M_{500}\gtrsim5\times10^{14}~\textrm{M}_{\odot}$) objects at $z\sim1$, both comparable to present SZ observational surveys in terms of mass and sufficient to derive robust conclusions. The deepest current observational survey follow-up data sets resolve the inner structure at the tens of kiloparsec scale \citep{Bartalucci2017}. A simulation reproducing these characteristics would require a $\gtrsim1~\textrm{Gpc}^3$ volume to be simulated at high resolution. This being impossible given current computational resources, we adopted a strategy in which the total volume was split into three periodic boxes of $1\,h^{-1}$ comoving Gpc on a side containing $2048^3$ dark matter particles, and then used the `zoom-in' technique to further refine individual systems from these parent large box simulations. The simulations are described in \cite{LeBrun2018}. In short, the simulations were carried out with the Eulerian adaptive mesh refinement (AMR) code \ramses~\citep{Teyssier2002} with cosmological parameters taken from \cite{Planck2015}. The boxes have a spatial resolution of $\sim15~h^{-1}$ comoving kpc (ckpc). Haloes were found using \textsc{phew}, recentred using a shrinking sphere procedure \citep{Power2003} and then spherical overdensity masses $M_{\Delta}$ were computed. The 'zoom-in' simulations have an effective resolution of $8192^3$ (8K) particles, resulting in a particle mass of $m_{cdm}=1.59\times10^8~h^{-1}~\textrm{M}_{\odot}$. The main haloes thus contain at least $\sim2$ million particles within $\rv$. Their spatial resolutions is $\sim3.8~h^{-1}$ ckpc.

Here we focus on the dark matter only (DMO) `zooms' of the 25 most massive systems selected in four different redshift slices ($z_{\rm sel}=1,~0.8,~0.6$ and 0\footnote{This redshift distribution has been chosen to match an observational sample at $z\gtrsim0.5$ \cite{Bartalucci2019} as well as providing a `local reference' for comparison with both observations and previous theoretical work.}). The main characteristics of the subsample at each $z_{sel}$, including the resulting minimum mass, are summarised in Table 1 of \cite{LeBrun2018}. A resolution study shows that the 8K zooms (i.e.\ the nominal resolution) are converged (in particular with the even-higher resolution 16K runs) over the whole resolved ($r\ge r_{\rm min}^{\rm 8K}$) radial range (see Appendix A of \cite{LeBrun2018}) and that an effective resolution of $8192^3$ is thus the minimum required. 

\section{Results}
\label{sec:res}

\begin{figure}
\centering
\sidecaption
\includegraphics[width=0.74\columnwidth]{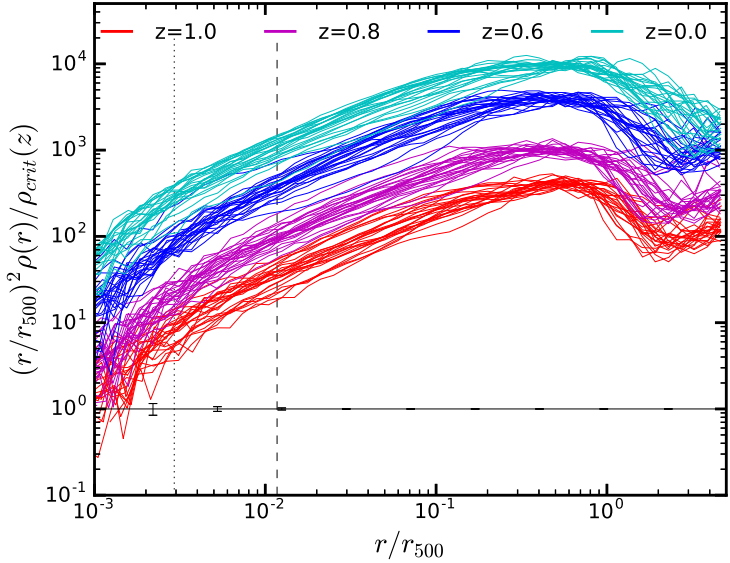}
\caption{Spherical density profiles for each redshift slice. For clarity, the profiles of the $z_{sel}=1,~0.8,~0.6$ and 0 systems have been shifted up by a factor of 2, 5, 20 and 50, respectively. The resolution limit at $z=0$ is depicted as dashed and dotted vertical lines for the boxes and the DMO zooms, respectively. The typical Poisson errors for every fifth data point are displayed as black error bars. Figure reproduced from \cite{LeBrun2018}.}
\label{fig:rhoprof}
\end{figure}

Figure~\ref{fig:rhoprof} depicts the density profiles computed using 50 evenly-spaced logarithmic bins over the $10^{-3}\leq r/\rv\leq5$ radial range with the \textsc{pymses}\footnote{http://irfu.cea.fr/Projets/PYMSES/intro.html} python module.  We use mass-weighted radii in  this and subsequent figures. Densities were normalised by the critical density of the Universe at the cluster redshift, and  the radii by the corresponding $\rv$.  The resolution limits at $z=0$ are shown as black dashed and dotted lines for the boxes and the DMO zooms, respectively. Both limits are nearly redshift-independent. The `zoom-in' approach allows for at least a five-fold increase of spatial resolution. Being smaller than the Poisson errors, fluctuations are real and plausibly due to substructures and oscillations resulting from halo relaxation. 

\begin{figure}[h]
\includegraphics[width=0.5\columnwidth]{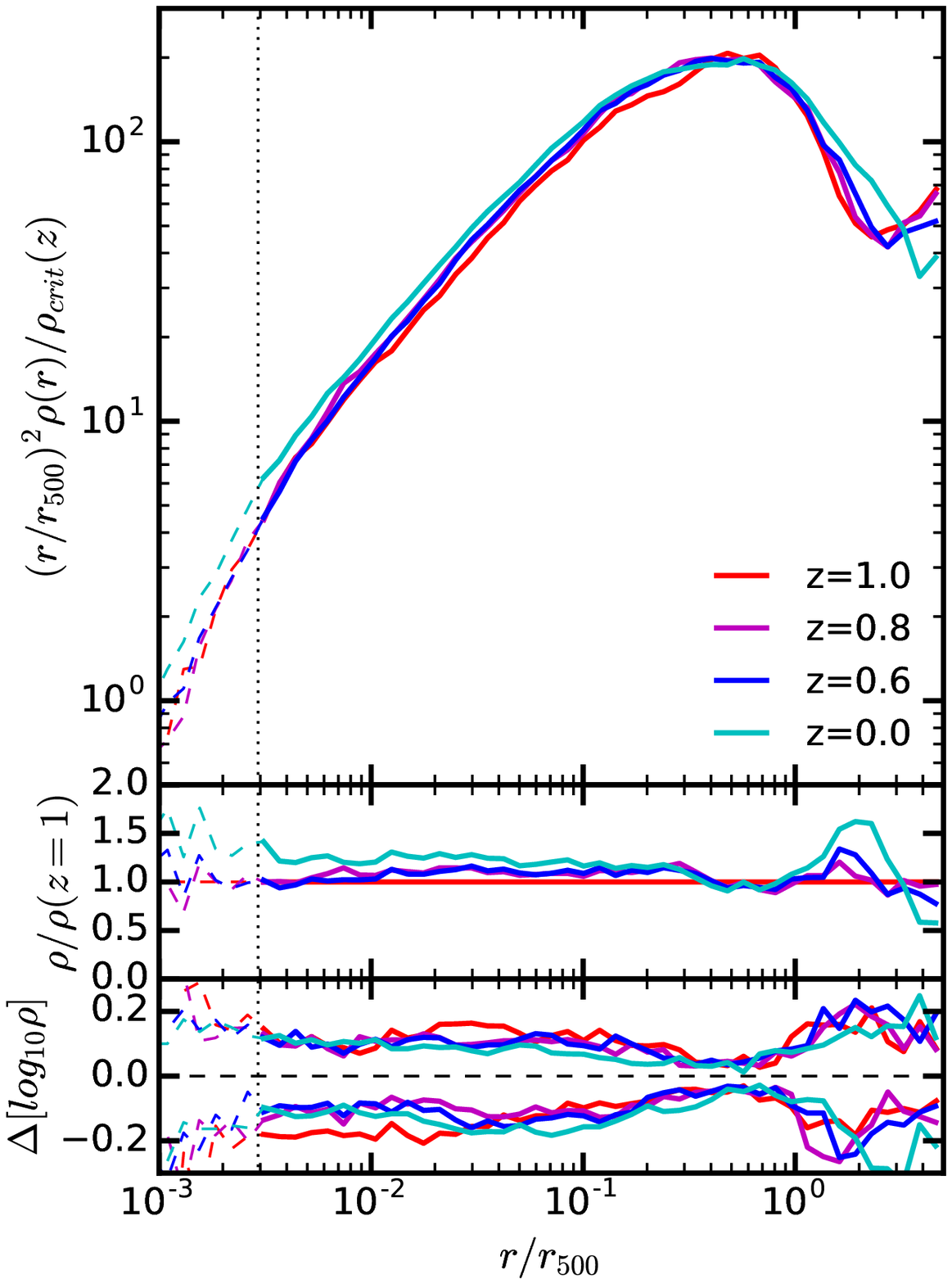}
\includegraphics[width=0.5\columnwidth]{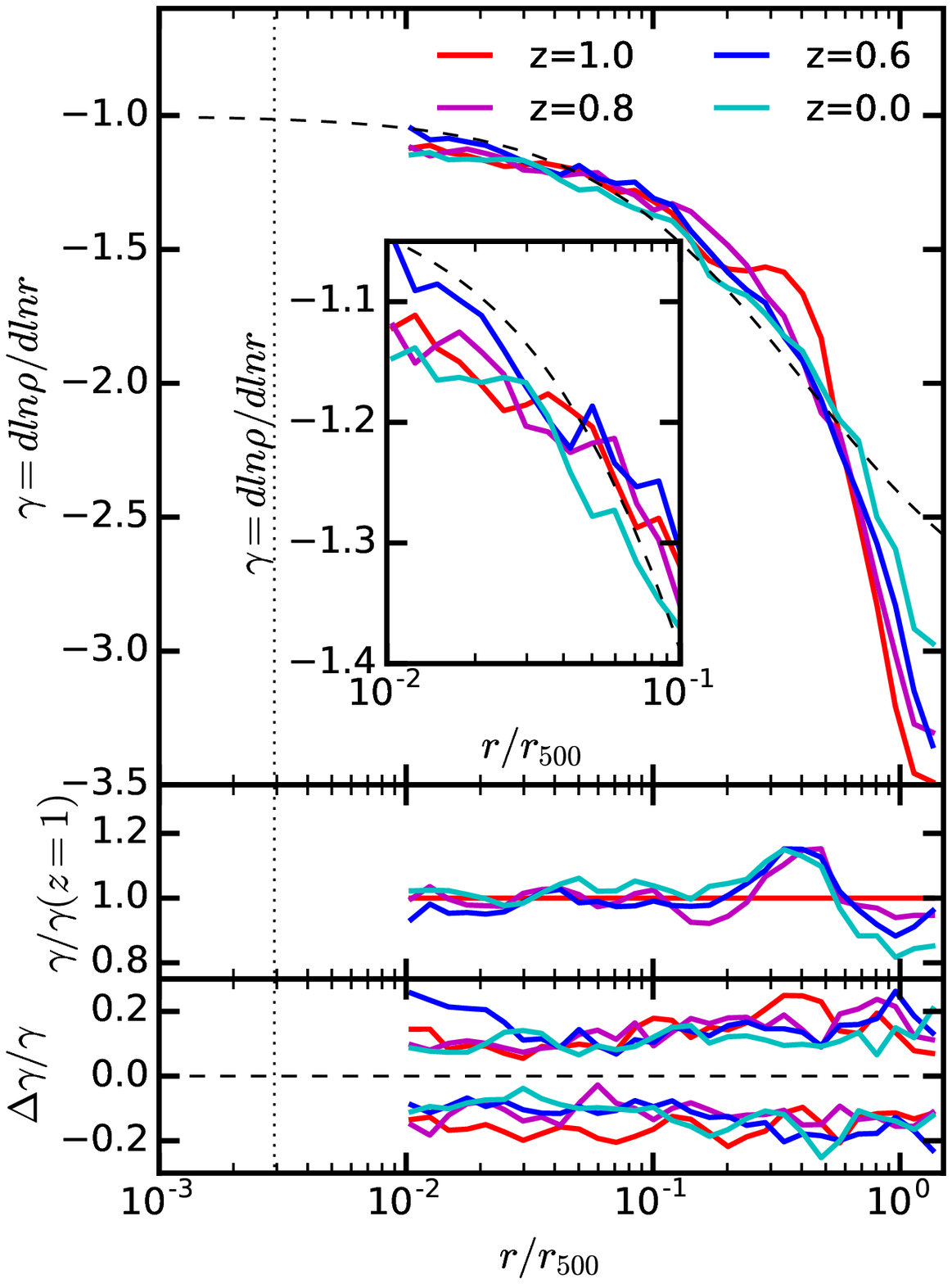}
\caption{Median spherical density (\emph{left}) and logarithmic slope (\emph{right}) profiles for each redshift slice (\emph{top}). The \emph{middle} panels show the median profiles normalised by the median profile at $z_{sel}=1$ whereas the \emph{bottom} panel displays the difference between the 16/84 percentiles and the median. In all panels, the resolution limit at $z=0$ is depicted as a dotted line and the profiles are plotted either as dashed thinner lines below that limit for the density or only from the eighth radial bin above that limit for the logarithmic slope where the Savitzky-Golay filter method is valid. The inset in the \emph{top right} panel is a zoom of the $0.01\le r/\rv\le0.1$ radial range. Figure reproduced from \cite{LeBrun2018}.}
\label{fig:rhoprofbis}
\end{figure}

The left panels of Fig.~\ref{fig:rhoprofbis} show the median scaled density profiles of each redshift slice. In the self-similar model, we expect the density profiles to be alike, regardless of cluster redshift or any other properties. Hereafter, evolution refers to the evolution of the scaled profiles, i.e.\ evolution in excess of the self-similar expectation. The mean density within  $\rv$ being  proportional to  $\rho_{\rm crit}(z)$ by definition,  scaled profiles  can solely have varying shapes in the region $r/\rv<1$.  Additionally, they are predicted to intersect at  $r/\rv\sim0.6$, the radius enclosing half of $M_{500}$,  and their scatter is hence smallest around that radius.  Beyond $\rv$,  a breach of self-similarity may turn  into evolution in both shape and normalisation. The scaled profiles display a barely noticeable evolution for $1\geq z\geq 0.6$ over the whole radial range. They change  by never more than a factor of $\sim1.2$ with decreasing redshift (see the \emph{middle} panel).  Conversely, they evolve a bit more (by up to a factor of $\sim1.5$) both in the core ($r/\rv\lesssim0.2$) and the outskirts ($r/\rv\gtrsim1.5$) for $z\leq0.6$. This is consistent with the `stable clustering' hypothesis often used for computing the non-linear matter power spectrum \citep[e.g.][]{Peebles1980}. The former evolution corresponds to a small increase of  peakiness. The latter is mostly due to their mass growth. The evolution in the outskirts corresponds to the transition between the one- and two-halo terms moving outwards as the haloes grow. The scatter of the density profiles, as displayed in the \emph{bottom} panel,  is notably small over the  redshift and radial ranges ($\lesssim0.2$ dex).  The scatter is less than $0.15$ dex within $\rv$. In  the outskirts, it is slightly larger  than in the core  and seems to increase mildly (by $\sim30$ per cent) with redshift. Note that these trends have to be confirmed with the whole set of simulations as each redshift slice only contains 25 systems.

The high degree of self-similarity is confirmed when studying the profile shapes. To quantify the profile shapes, we calculated the logarithmic slope $\gamma\equiv \textrm{d}\ln\rho/\textrm{d}\ln r$ from the smallest resolved radius ($\max(r_{min}/\rv)\lesssim0.003)$,  using the fourth-order Savitzky-Golay algorithm over the 15 nearest bins \mbox{\citep{Savitzky1964}}. Excluding the seven innermost and outermost bins where the method is no longer valid, the logarithmic slope profiles cover  the $0.01\leq r/\rv \leq 1.35$ radial range. The right panels of Fig.~\ref{fig:rhoprofbis} show the median logarithmic slope profiles of each redshift slice. They display small amounts of both evolution (it never exceeds 20 per cent; see the \emph{middle} panel) and dispersion ($\lesssim0.2$ dex; see the \emph{bottom} panel) over the whole redshift and radial range. Similarly to the density profiles, the evolution is more important in the very central  regions ($r/\rv\lesssim0.02$) and the outer regions ($r/\rv\gtrsim0.6$).  More surprising, the slope of the profiles, contrarily to the profiles themselves, displays noticeable evolution for all $z\leq1$ and not just  $z\leq0.6$. The scatter of the density slope exhibits nearly no evolution over the whole radial range and it has similar amplitude to that of the profiles (except around $r/\rv\sim0.6$, where the scatter of the density is minimum by construction).  Crucially, the inner slope presents no signs of converging to an asymptotic value but is still consistent with a NFW profile of typical concentration for massive galaxy clusters ($c_{200}=3.5$) over the radial range in question (see the \emph{top} panel and its inset; in agreement with the results of e.g.\ \cite{Navarro2004}). 

\begin{figure}
\centering
\sidecaption
\includegraphics[width=0.74\columnwidth]{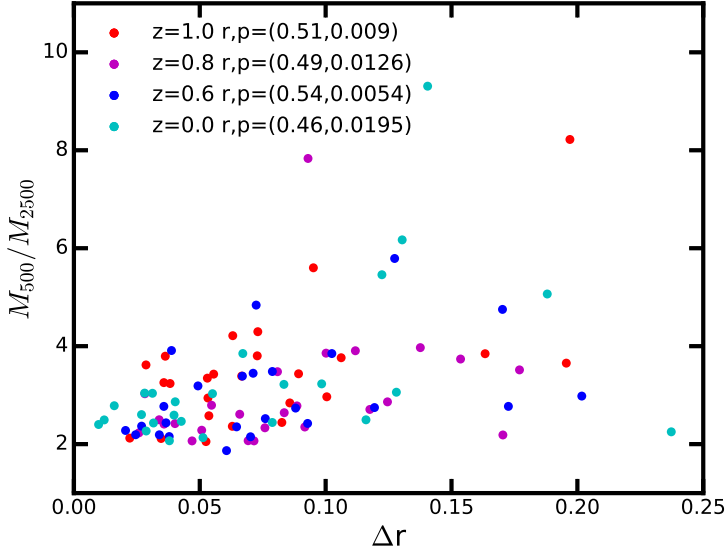}
\caption{$M_{500}/M_{2500}$ as a function of relaxation state for each redshift slice. The Spearman's rank and the null hypothesis probability are listed in the legend. Figure reproduced from \cite{LeBrun2018}.}
\label{fig:sparsity}
\end{figure}

A first investigation of the origin of the scatter in the density profiles is illustrated by Fig.~\ref{fig:sparsity}, which depicts the $M_{500}/M_{2500}$ ratio as a function of $\Delta r$\footnote{$\Delta r$ is the distance between the centre of mass within the \cite{Bryan1998} virial radius $r_{vir}$ and the centre of the shrinking sphere, in units of $r_{vir}$.}. The most relaxed clusters ($\Delta r\leq0.04$) lean towards being more centrally concentrated ($M_{500}/M_{2500}\lesssim4$). Yet,  there is no clear correlation between both parameters   in terms of Spearman's rank and null hypothesis probability,  as listed in the legend of Fig.~\ref{fig:sparsity}. In fact, the distribution is consistent with all systems having a similar $M_{500}/M_{2500}$ ratio, no matter their relaxation state, but with a dispersion that rises with rising $\Delta r$. Hence, the scatter in the density profiles is only connected to the relaxation state of the galaxy cluster  through the fact that relaxed clusters are mostly centrally concentrated, while  unrelaxed objects span a larger variety of profile shapes, including very shallow profiles. Note that the entire simulation sample is needed to reach a reliable conclusion.

\section{Discussion and conclusion}
\label{sec:sum}

We have focussed in this work on the dark matter density profiles of the 25 most massive systems selected from DMO simulations in four redshift slices ($z_{\rm sel}=1,~0.8,~0.6$ and 0).  With a median mass of $M_{500} = 6.3\times10^{14} M_{\odot}$ at $z=1$, the sample is composed of the rarest objects, probing for the first time the extreme limits of the cluster mass function, such as would be detectable observationally only in all-sky surveys. Surprisingly, these objects exhibit a high level of self-similarity, and their dark matter density profiles can be described with the typical NFW profile found in relaxed local systems. 

From the Millenium simulations \citep*{Fakhouri2010},  high--mass systems with  $M>5\times 10^{14}~\textrm{M}_{\odot}$  at $z=1$ ($z=0$)  have undergone at least one major merger (mass ratio $>1:3$) during the preceding 4 Gyr (12 Gyr). The relaxation time estimated from their crossing time ($t_{cross}\propto \rv/\sigma_{500}$ with $\sigma_{500}=(GM_{500}/\rv)^{1/2}$) is close to two Hubble times ($t_{H}=1/H(z)$), i.e.\ about 16 Gyr (29 Gyr). A similar conclusion is reached  if one uses the dynamical time. We expect that these objects should still be forming and thus be highly unrelaxed. Using $\Delta$r  as a dynamical indicator, or from visual inspection of the images, we find that the vast majority of the systems in question are indeed unrelaxed. Naively, one would then expect that the  density profiles of such objects would exhibit large variations, linked to the wide variety of dynamical states and formation histories, and that the median profile has not yet converged to the near-universal form of relaxed systems in the local Universe. In contrast,  once scaled according to the critical density at each redshift, the density profiles of the clusters in our simulations are remarkably similar, with a low dispersion of less than 0.15 dex within $\rv$. Furthermore, there is little evolution in  the radial logarithmic slope or scatter with redshift. This surprising result suggests that the `universal', `broken/running' power-law, density profile (e.g.\ similar to NFW or Einasto)  is already in place at $z>1$ and that it is robust to merging activity. This conclusion is similar to that recently  obtained   for primordial (Earth--mass)  haloes  by \cite{Angulo2016} and \cite{Ogiya2016}, but at scales that are 21 orders of magnitude larger. Interestingly, \cite{McDonald2017}  recently found  a remarkably standard self-similar evolution  in the mean profile of the hot gas beyond  the cooling core region in massive clusters up to $z\sim1.9$.  This would be  a natural consequence of the self-similar evolution of the underlying dark matter distribution that we have shown here, since  these systems are dark-matter dominated and the gas evolution, except in the very central regions,  is dominated by simple gravitational  physics. There is an indication of a residual link between the profile shape and dynamical state, with the most unrelaxed clusters exhibiting a larger dispersion.   Ongoing work on a larger sample, covering a wider mass range, will enable us to better characterise both the evolution and the scatter. These are essential to understand the link between profile shape, dynamical state, and formation history. This new study should thus start bringing value to observational analysis in the near-future. Its results will be made available to current and future experiments and will be extremely useful to understand the impact of dynamical state (e.g. due to recent mergers) on mass measurements and profile shapes as well as for obtaining unbiased mass profiles. A precise quantification of the impact of cosmology on profile shapes is still badly needed and is the topic of a lot of ongoing studies.

\section*{Acknowledgements}
This work was supported by the French Agence Nationale de la Recherche under grant ANR-11-BS56-015 and by the European Research Council under the European Union's Seventh Framework Programme (FP7/2007-2013) / ERC grant agreement number 340519. It was granted access to the HPC resources of CINES under the allocations 2015-047350, 2016-047350 and 2017-047350 made by GENCI. The authors also thank the anonymous referee for useful suggestions that helped improve the paper. AMCLB is extremely grateful to Damien Chapon for his help with \textsc{pymses} and Patrick Hennebelle for granting access to his shared memory machine. She further thanks for its hospitality the Institute for Computational Science of the University of Z\"urich.

\bibliography{25mostmassive.bib}

\end{document}